\documentclass[aps,prd,showpacs,nofootinbib,superscriptaddress,11pt]{revtex4}  

\usepackage{graphicx}
\usepackage{epstopdf}
\usepackage{amsmath}
\usepackage{amsfonts}
\usepackage{amssymb}
\usepackage{appendix}
\usepackage{comment}
\usepackage{bbold}
\usepackage{color}
\usepackage{slashed}
\usepackage{subfigure}
\usepackage{setspace}
\usepackage{footnote}
\usepackage{multirow}
\usepackage[normalem]{ulem}

\begin{document}

\singlespacing

\title{Large, Extra Dimensions at the Deep Underground Neutrino Experiment}

\author{Jeffrey M. Berryman}
\affiliation{Northwestern University, Department of Physics \& Astronomy, 2145 Sheridan Road, Evanston, IL 60208, USA}
\author{Andr\'{e} de Gouv\^{e}a} 
\affiliation{Northwestern University, Department of Physics \& Astronomy, 2145 Sheridan Road, Evanston, IL 60208, USA}
\author{Kevin J. Kelly}
\affiliation{Northwestern University, Department of Physics \& Astronomy, 2145 Sheridan Road, Evanston, IL 60208, USA}
\author{O. L. G. Peres}
\affiliation{~Instituto de F\'isica Gleb Wataghin - UNICAMP, {13083-859}, Campinas SP, Brazil}
\affiliation{~The Abdus Salam International Centre for Theoretical Physics, Strada Costiera 11, 34014 Trieste, Italy.}
\author{Zahra  Tabrizi}
\affiliation{Departamento de F\'{\i}sica Matem\'atica, Instituto de F\'{\i}sica,
 Universidade de S\~ao Paulo, C.\ P.\ 66.318, 05315-970 S\~ao Paulo, Brazil
}
\affiliation{School of Particles and Accelerators, Institute for Research in Fundamental Sciences (IPM), P.O.Box 19395-1795, Tehran, Iran}


\begin{abstract}
We investigate the potential of the long-baseline Deep Underground Neutrino Experiment (DUNE) to study large-extra-dimension (LED) models originally proposed to explain the smallness of neutrino masses by postulating that right-handed neutrinos, unlike all standard model fermion fields, can propagate in the bulk. The massive Kaluza-Klein (KK) modes of the right-handed neutrino fields modify the neutrino oscillation probabilities and can hence affect their propagation. We show that, as far as DUNE is concerned, the LED model is indistinguishable from a $(3+3N)$-neutrino framework for modest values of $N$; $N=1$ is usually a very good approximation. Nonetheless, there are no new sources of $CP$-invariance violation other than one $CP$-odd phase that can be easily mapped onto the $CP$-odd phase in the standard three-neutrino paradigm. We analyze the sensitivity of DUNE to the LED framework, and explore the capability of DUNE to differentiate the LED model from the three-neutrino scenario and from a generic $(3+1)$-neutrino model. 

\end{abstract}


\maketitle


\section{Introduction}
\label{Introduction}
Neutrino oscillation experiments have revolutionized our understanding of the neutrino sector of the standard model (SM). It is now established that at least two of the three known neutrinos are massive, and that the mass and flavor eigenstates are distinct. There are still several unanswered questions in neutrino physics, including the neutrino mass hierarchy, the potential existence of new neutrino states, and the status of $CP$ invariance in the lepton sector. To address these questions and further investigate the neutrino oscillation phenomenon, we need a new generation of neutrino oscillation experiments. The long-baseline Deep Underground Neutrino Experiment (DUNE) in the U.S.~\cite{Acciarri:2016crz,Acciarri:2015uup} and the Hyper-Kamiokande (HyperK) experiment in Japan~\cite{Abe:2015zbg} are proposed to answer these and several other questions, and are poised to provide qualitatively better and more precise tests of the current three-massive-neutrinos paradigm.  
 
 Although the absolute neutrino masses are not yet determined, we can indirectly infer from cosmic surveys that the known neutrino masses are below the eV-scale~\cite{Ade:2015xua}. Similar bounds, albeit weaker but more direct, come from kinematical probes of nonzero neutrino masses \cite{Kraus:2004zw,Aseev:2011dq}. The fact that neutrino masses are much smaller than all known fermion masses in the SM is widely interpreted as evidence that the mechanism behind neutrino masses is different from that of all other known particles. The hypothesis that there are more, compactified dimensions of space, and that these are large (i.e.,  much larger than the inverse of the Planck mass) was introduced in order to address the infamous SM hierarchy problem \cite{ArkaniHamed:1998rs,ArkaniHamed:1998nn,Antoniadis:1998ig}, and also provides a mechanism for understanding why neutrino masses are parametrically smaller than charged-fermion masses. In these large-extra-dimension (LED) models, it is natural to assume that singlets of the SM gauge group, such as the graviton or the right-handed neutrino states,  can propagate unconstrained in all dimensions, while the SM-charged objects are confined to a four-dimensional spacetime. If there are right-handed neutrino fields that propagate in the bulk (or a subset of the bulk), the equivalent four-dimensional neutrino Yukawa couplings are suppressed relative to charged-fermion Yukawa couplings by a factor proportional to the volume of the extra dimensions~\cite{Dienes:1998sb,ArkaniHamed:1998vp}. In these scenarios, neutrinos are very light for the same reason gravity appears to be very weakly coupled. 
 
 The Kaluza-Klein (KK) modes of the higher-dimensional right-handed neutrino fields behave as an infinite tower of sterile neutrinos. If these are light enough, one expects deviations from the three-massive-neutrinos paradigm in neutrino oscillation experiments. The neutrino oscillation phenomenology of LED models has been extensively studied in the literature (see, for example, Refs.~\cite{Dvali:1999cn,Mohapatra:1999zd,Davoudiasl:2002fq,Machado:2011jt,BastoGonzalez:2012me,Rodejohann:2014eka,DiIura:2014csa}). It has also been proposed~\cite{Machado:2011kt} that the reactor anomaly can be explained within the LED framework.  More generically, the equivalence between the LED model and a framework with several sterile neutrinos was discussed in~\cite{Esmaili:2014esa}. Other phenomenological aspects of LED models and their application to nonzero neutrino masses have also been explored in depth in the literature (see for example, Refs.~\cite{deGouvea:2001mz,Bhattacharyya:2002vf}).
 
We study the potential of the Deep Underground Neutrino Experiment (DUNE) to exclude or observe the effects of the LED model, and investigate how well DUNE can constrain the LED parameters. Highlights include the discussion of $CP$-invariance violation phenomena in the LED model using the DUNE experiment. Several other new physics scenarios can be studied using the  precise measurements of the DUNE experiment. The capability of DUNE to test the one-sterile neutrino hypothesis was recently explored in detail in Ref.~\cite{Berryman:2015nua,Gandhi:2015xza} while the effects of non-standard interactions (NSI) of neutrinos were investigated in~\cite{Masud:2015xva,Coloma:2015kiu,deGouvea:2015ndi,Liao:2016hsa}. Here, we also explore the ability of DUNE to differentiate the LED hypothesis from the three-neutrino and the four-neutrino hypotheses.

The paper is organized as follows: We discuss the LED formalism and the related neutrino oscillation probabilities in Section~\ref{Formalism}. The sensitivity of DUNE to the LED hypothesis is studied in Section~\ref{Sensitivity}, and we demonstrate the capability of DUNE to measure non-zero LED parameters in Section~\ref{Measuring}. Section~\ref{Diagnosis} is devoted to studying the ability of DUNE to differentiate qualitatively distinct scenarios. We summarize our results and offer some conclusions in Section~\ref{Concluding}.
 

\section{Formalism and Oscillation Probabilities}
\label{Formalism}
In this section we discuss the neutrino oscillation probabilities in LED models, and restrict our discussion to models with one relevant extra-dimension. We extend the SM with three massless five-dimensional gauge singlet fermions $\Psi^{\alpha}\equiv (\psi_L^{\alpha},\psi_R^{\alpha})$ associated to the three active neutrinos $\nu_L^\alpha$. The indices $\alpha$ correspond to $e,\mu,\tau$, in spite of the fact that there are no charged leptons associated to $\Psi^{\alpha}$. The fifth dimension is compactified with periodic boundary conditions in such a way that, from a four-dimensional point of view,  $\Psi^{\alpha}$ can be decomposed into a tower of Kaluza-Klein (KK) states $\psi^{(n)}_{L,R}~(n=0,\pm1,\cdots,\pm\infty)$. Redefining the new fields as $\nu^{\alpha(0)}_R\equiv\psi^{\alpha(0)}_R$ and $\nu^{\alpha(n)}_{L,R}\equiv\Big(\psi^{\alpha(n)}_{L,R}+\psi^{\alpha(-n)}_{L,R}\Big)/\sqrt 2,~(n=1,...,\infty)$, the mass terms of the Lagrangian, after electroweak symmetry breaking, are~\cite{ArkaniHamed:1998vp, Dienes:1998sb, Barbieri:2000mg}:
\begin{eqnarray}\label{massterms}
\mathcal{L_{\rm{mass}}}&=&m^D_{\alpha\beta}(\bar\nu^{\alpha(0)}_R\nu^{\beta}_L+\sqrt{2}\sum^{\infty}_{n=1}\bar\nu^{\alpha(n)}_R\nu^{\beta}_L)+\sum^{\infty}_{n=1}\frac{n}{R_{\rm{ED}}}\bar\nu^{\alpha(n)}_R\nu^{\alpha(n)}_L+\rm{h.c.}, \nonumber\\
&\equiv&\sum_{i=1}^3\bar{\mathcal{N}}^i_RM^i\mathcal{N}^i_L+\rm{h.c.},
\end{eqnarray}
where $m^D$ is the Dirac mass matrix proportional to the neutrino Yukawa couplings and $R_{\rm{ED}}$ is the radius of compactification. Note that all massive fermions are Dirac fermions. It is convenient to define pseudo mass eigenstates $\mathcal{N}^i_{L(R)}$ by rotating the neutrino states to a basis in which $m^D$ is diagonal:
\begin{eqnarray}\label{pseudo_basis}
{\mathcal{N}^i_{L(R)}}=\Big(\nu^{i(0)},\nu^{i(1)},\nu^{i(2)},\cdots\Big)^T_{L(R)},  ~~~~~{\rm{and}}~~~~~ M^i=
\begin{pmatrix}
m_i^D&0&0&0&\ldots\\
\sqrt{2}m_i^D&1/R_{\rm{ED}}&0&0&\ldots\\
\sqrt{2}m_i^D&0&2/R_{\rm{ED}}&0&\ldots\\
\vdots&\vdots&\vdots&\vdots&\ddots
\end{pmatrix},
\end{eqnarray}
where $m_i^D$ are the elements of the diagonalized Dirac mass matrix $(m^D)_d=\mathrm{diag}(m^D_1,m^D_2,m^D_3)$. 
The relation between the active neutrinos in the SM and the corresponding pseudo mass eigenstates is given by 
\begin{eqnarray}\label{flavor_pseudo}
\nu^{\alpha}_L=\sum_{i=1}^3U^{\alpha i}\nu^{i(0)}_L, ~~~~~(\alpha=e,\mu,\tau),
\end{eqnarray}
where the $3\times3$ unitary matrix $U$ describes the mismatch between the flavor and pseudo mass eigenstates of neutrinos. This matrix is parametrized by three mixing angles $(\theta_{12},\theta_{13},\theta_{23})$ and one Dirac $CP$-violating phase $\delta_{13}$. In the limit where $m^D\times R_{\rm ED}\to 0$, the KK modes and the active neutrinos decouple, and $U$ is the standard neutrino mixing matrix for Dirac neutrinos. We are interested in values of $R_{\rm ED}$ such that  $R^{-1}_{\rm ED}$ is larger than $m^D_i$, but small enough that nontrivial effects might be observed in long-baseline oscillation experiments.

The true neutrino masses are found by diagonalizing the $n\times n$ matrix $M_i^\dagger M_i$ with an $n\times n$ unitary matrix $S$ as: $S_i^{\dagger}M_i^{\dagger}M_iS_i$. Therefore, the true mass eigenstates are ${\mathcal{N}_i^\prime}_L=\Big(\nu^{\prime(0)}_i,\nu^{\prime(1)}_i,\nu^{\prime(2)}_i,\cdots\Big)^T_{L}=S_i^{\dagger}{\mathcal{N}_i}_L$. Using Eq.~(\ref{flavor_pseudo}) we can obtain a relation between the active neutrinos of the SM and the mass eigenstates of the KK neutrinos,
\begin{equation}\label{mass_flavor}
{\nu_{\alpha}}_L=\sum_{i=1}^3U_{\alpha i}{\nu_{iL}^{(0)}}=\sum_{i=1}^3U_{\alpha i}\sum_{n=0}^{\infty}S_i^{0n}{\nu_{iL}^{\prime(n)}}~,~~~~~~~(\alpha=e,\mu,\tau),
\end{equation}
where 
\begin{equation}\label{Si0n}
\left(S^{0n}_i\right)^2=\frac{2}{1+\pi^2\left(m_i^{D}R_{\rm{ED}}\right)^2+\left(\lambda^{(n)}_i\right)^2/\left(m_i^{D}R_{\rm{ED}}\right)^2}. 
\end{equation}
Above, $({\lambda_i^{(n)}})^2$ are the eigenvalues of the matrices $R^2_{\rm{ED}}M_i^{\dagger}M_i$, and are obtained by solving the following transcendental equation~\cite{Dvali:1999cn,Dienes:1998sb,Barbieri:2000mg}: 
\begin{equation}\label{trans}
\lambda_i^{(n)}-\pi \big(m_i^DR_{\rm{ED}}\big)^2\cot\left(\pi\lambda_i^{(n)}\right)=0.
\end{equation}
The roots of this transcendental equation satisfy the relation $n\le\lambda_i^{(n)}\le (n+1/2)$, so the masses of the neutrino states in the LED model are
\begin{equation}\label{masses}
m_i^{(n)}=\frac{\lambda_i^{(n)}}{R_{\rm{ED}}}\simeq \frac{n}{R_{\rm{ED}}},~~~~(n=0,1,\cdots),
\end{equation}
where $n=0$ and $n\ge1$ correspond to the mostly active and mostly sterile neutrinos, respectively. As mentioned earlier, we are interested in $R^{-1}_{\rm ED}\gg m^D$. 

The Dirac masses $(m_1^D,m_2^D,m_3^D)$ which appear in the Hamiltonian are not the masses of the mostly active neutrinos. They are, however, related to the mostly active neutrino masses and are hence constrained by neutrino oscillation data, along with $R_{\rm ED}$. The solar and atmospheric mass-squared differences are 
\begin{eqnarray}\label{mass_relation}
\Delta m^2_{\rm{sol}}\equiv\Delta m^2_{21}=\frac{\left(\lambda_2^{(0)}\right)^2-\left(\lambda_1^{(0)}\right)^2}{R_{\rm{ED}}^2},~~~{\rm{and}}~~~\Delta m^2_{\rm{atm}}\equiv|\Delta m^2_{31}|=\Big|\frac{\left(\lambda_3^{(0)}\right)^2-\left(\lambda_1^{(0)}\right)^2}{R_{\rm{ED}}^2}\Big|.
\end{eqnarray}
We can solve the equations above and replace two among $(m_1^D,m_2^D,m_3^D,R_{\rm ED})$ with $\Delta m^2_{21}$ and $\Delta m^2_{31}$, which are constrained by experiment.\footnote{We follow the discussion in~\cite{BastoGonzalez:2012me}. Explicitly, for the normal hierarchy (NH) case ($\lambda_1^{(0)}<\lambda_2^{(0)}<\lambda_3^{(0)}$), we use Eq.~(\ref{trans}) to find $\lambda_1^{(0)}$ as a function of $(m_1^D,R_{\rm{ED}})$ while Eq.~(\ref{mass_relation}) is used to express $\lambda_{2(3)}^{(0)}$ as a function of $\lambda_1^{(0)}$. Eq.~(\ref{trans}) then provides a relation between  $m_{2(3)}^D$ and $(m_1^D,R_{\rm{ED}})$. For the inverted hierarchy (IH) case ($\lambda_3^{(0)}<\lambda_2^{(0)}<\lambda_1^{(0)}$) we follow the same procedure to express $m_{1(2)}^D$ as a function of $(m_3^D,R_{\rm{ED}})$. Note that the equations above only have solutions for $0\le\lambda_i^{(0)}\le0.5$.} Hence, the LED framework can be characterized by the standard oscillation parameters -- $\theta_{12},\theta_{13},\theta_{23},\delta_{13}$, $\Delta m^2_{21}$, and $\Delta m^2_{31}$ -- and two new free parameters, which we choose to be  $m_0\equiv m^D_{1(3)}$ and $R_{\rm{ED}}$, for the NH (IH) case. 

Neutrino flavor evolution in the LED model is governed by the  following equation~\cite{Esmaili:2014esa}:
\begin{eqnarray}\label{evolution}
i\frac{d}{dr}{\mathcal{N}_i}_L=\Bigg[\frac{1}{2E_\nu}M_i^{\dagger}M_i{\mathcal{N}_i}_L+\sum_{j=1}^3
\begin{pmatrix}
\mathcal{V}_{ij} & 0_{1\times n} \\
0_{n\times 1} & 0_{n\times n}
\end{pmatrix}
{\mathcal{N}_j}_L\Bigg]_{n\to\infty},~\mathcal{V}_{ij}=\sum_{\alpha=e,\mu,\tau} U^*_{\alpha i}U_{\alpha j}\Big(\delta_{\alpha e}V_{\rm{CC}}+V_{\rm{NC}}\Big),\nonumber\\
\end{eqnarray}
where $V_{\rm{CC}}=\sqrt{2}G_FN_e$ and $V_{\rm{NC}}=-\sqrt{2}/2G_FN_n$ are the charged- and neutral-current matter potentials, $G_F$ is the Fermi constant and $N_{e(n)}$ is the electron (neutron) number density along the trajectory of the neutrinos. For the purposes of this manuscript, we assume the electron and neutron number densities to be the same and constant. As usual, $U_{\alpha i}\leftrightarrow U_{\alpha i}^*$ and the sign of the matter potentials are reversed when one considers the flavor evolution of antineutrinos. 

The equivalence between the LED model and a $(3+3N)$ sterile framework with $N$ KK modes was explored in detail in Ref.~\cite{Esmaili:2014esa}. The flavor and mass eigenstates in a $(3+3N)$ framework are related by a $(3+3N)\times(3+3N)$ unitary matrix $W$,
\begin{eqnarray}\label{mass_flavor2}
{\mathcal{N}_\alpha}_L=\sum_{l=1}^{3+3N}W_{\alpha l}{{\mathcal{N}_l^\prime}}_L,
\end{eqnarray}
where ${\mathcal{N}_\alpha}_L=\Big(\nu_e,\nu_\mu,\nu_\tau,\nu_{s_1},\nu_{s_2},\nu_{s_3},\cdots\Big)^T_L$, in which $\nu_{s_i}$ are the sterile eigenstates. Comparing Eqs.~(\ref{mass_flavor}) and (\ref{mass_flavor2}), 
\begin{eqnarray}
{\nu_\alpha}_L=\sum_{i=1}^3U_{\alpha i}\sum_{n=0}^NS_{i}^{0n}{\nu_{iL}^{\prime(n)}}=\sum_{i=1}^3 \sum_{n=0}^N W_{\alpha (i+3n)}{\nu_{iL}^{\prime(n)}},~~~~(\alpha=e,\mu,\tau),
\end{eqnarray}
so
\begin{eqnarray}\label{Wmatrix}
 W_{\alpha (i+3n)}&=&U_{\alpha i}S_i^{0n},~~~~~~(i=1,2,3), ~~~~~(\alpha=e,\mu,\tau),~~~~~(n=0,1,\cdots,N).
\end{eqnarray}
For $R^{-1}_{\rm ED}\gg m^D$ we have $|S_i^{0n}|^2\propto n^{-2}$, so KK modes slowly decouple as they get heavier. This implies that there is a finite value of $N$ above which the $3+3N$ model is indistinguishable from the LED model. In practice, we have considered 2 KK modes in our calculations and have verified that the inclusion of more KK modes does not change our results. In fact, we have verified that, for the simulations performed here, 1 KK mode is sufficient. We further justify this approximation below. 
    
When matter effects can be ignored, the oscillation probabilities are
\begin{eqnarray}\label{eq5}
P (\nu_\alpha\to\nu_\beta)=\delta_{\alpha\beta} &-&4 \sum_{l>m}\Re \big[W_{\alpha l}W^*_{\beta l}W^*_{\alpha m}W_{\beta m}\big]\sin^2\Big(\frac{\Delta m^2_{lm}L}{4E_\nu}\Big)\nonumber\\
&+&2 \sum_{l>m}\Im \big[W_{\alpha l}W^*_{\beta l}W^*_{\alpha m}W_{\beta m}\big]\sin\Big(\frac{\Delta m^2_{lm}L}{2E_\nu}\Big),~~~~(l,m=1,\cdots,3+3N),\nonumber\\
\end{eqnarray}
where $L$ is the oscillation baseline, $E_\nu$ is the neutrino energy, and $\Delta m^2_{lm}\equiv m_l^2-m_m^2$ with $m_{l=i+3n}\equiv m_i^{(n)}=\frac{\lambda_i^{(n)}}{R_{\rm{ED}}}$. Matter effects will modify the oscillation probabilities in a well-known way.\footnote{Matter effects can lead to resonant flavor-conversion. For the effective two-neutrino system $\nu_i^{\prime(n)}-\nu_i^{\prime(0)}$ in the LED model, the resonance condition occurs for very high neutrino energies ~\cite{Nunokawa:2003ep,Esmaili:2014esa}:
\begin{equation}
\label{resonance}
E_\nu^{\rm{res}}=\frac{\left(\lambda_i^{(n)}\right)^2-\left(\lambda_i^{(0)}\right)^2}{2V_{\rm {NC}} R_{\rm{ED}}^2}\simeq \frac{n^2}{2V_{\rm {NC}}}\frac{R_{\rm ED}^{-1}}{2~\rm eV}\simeq n^2 ~{\rm{TeV}}. 
\end{equation}
We are interested in the DUNE experiment, where neutrino energies are of order 1~GeV, and hence do not need to worry about the the resonant conversion of the active states into sterile KK modes.}

$CP$-invariance violation in the neutrino sector manifests itself as an asymmetry between the oscillation probabilities of neutrinos and antineutrinos. In the three-neutrino scenario, the only source of $CP$ violation (if the neutrinos are Dirac particles, which is the case here) is the phase $\delta_{13}$ in the leptonic mixing matrix $U$. In a generic $3+3N$ massive Dirac neutrinos framework there are $(3N+2)(3N+1)/2 - 1$ $CP$-odd phases beyond $\delta_{13}$ associated to the $(3+3N)\times(3+3N)$ unitary mixing matrix. In the LED model, however, the  $(3+3N)\times(3+3N)$ unitary matrix $W$ is not generic. As we can see from Eq.~(\ref{Wmatrix}), all the elements of $W$ are proportional to $U_{\alpha i}$ so all the new $CP$ phases are functions of $\delta_{13}$. Hence, while many of the elements of $W$ have nonzero $CP$-odd phases, no new $CP$-violating parameters are introduced within the LED framework. In other words, $CP$-violating phenomena are governed by the higher-dimensional neutrino Yukawa couplings, which define a $3\times 3$ matrix. This is identical to the familiar four-dimensional case when the neutrinos are Dirac fermions.  

For illustrative purposes, we evaluate the $S$ matrix numerically for $R_{\rm ED} = 5\times 10^{-5}~{\rm cm} = (0.38~\rm eV)^{-1}$ and $m_0=5\times 10^{-2}$~eV. The corresponding neutrino mixing matrix $W$ is, for the NH and IH, respectively,
\begin{eqnarray}
\label{W_num}
W_{\alpha i}^{\rm (NH)} &=
\left(\begin{array}{c c c c c c c c c c}
0.97 U_{e1} & 0.97 U_{e2} & 0.94 U_{e3} & 0.18 U_{e1} & 0.19 U_{e2} & 0.27 U_{e3} & 0.09 U_{e1} & 0.09 U_{e2} & 0.14 U_{e3} & \ldots\\
0.97 U_{\mu 1} & 0.97 U_{\mu 2} & 0.94 U_{\mu 3} & 0.18 U_{\mu 1} & 0.19 U_{\mu 2} & 0.27 U_{\mu 3} & 0.09 U_{\mu 1} & 0.09 U_{\mu 2} & 0.14 U_{\mu 3} & \ldots\\
0.97 U_{\tau 1} & 0.97 U_{\tau 2} & 0.94 U_{\tau 3} & 0.18 U_{\tau 1} & 0.19 U_{\tau 2} & 0.27 U_{\tau 3} & 0.09 U_{\tau 1} & 0.09 U_{\tau 2} & 0.14 U_{\tau 3} & \ldots\\
\vdots & \vdots & \vdots & \vdots & \vdots & \vdots & \vdots & \vdots & \vdots & \ddots\end{array}\right), \nonumber \\ \nonumber \\
W_{\alpha i}^{\rm (IH)} &=
\left(\begin{array}{c c c c c c c c c c}
0.95 U_{e1} & 0.94 U_{e2} & 0.97 U_{e3} & 0.26 U_{e1} & 0.27 U_{e2} & 0.18 U_{e3} & 0.13 U_{e1} & 0.14 U_{e2} & 0.09 U_{e3} & \ldots\\
0.95 U_{\mu 1} & 0.94 U_{\mu 2} & 0.97 U_{\mu 3} & 0.26 U_{\mu 1} & 0.27 U_{\mu 2} & 0.18 U_{\mu 3} & 0.13 U_{\mu 1} & 0.14 U_{\mu 2} & 0.09 U_{\mu 3} & \ldots\\
0.95 U_{\tau 1} & 0.94 U_{\tau 2} & 0.97 U_{\tau 3} & 0.26 U_{\tau 1} & 0.27 U_{\tau 2} & 0.18 U_{\tau 3} & 0.13 U_{\tau 1} & 0.14 U_{\tau 2} & 0.09 U_{\tau 3} & \ldots\\
\vdots & \vdots & \vdots & \vdots & \vdots & \vdots & \vdots & \vdots & \vdots & \ddots\end{array}\right), \nonumber \\
\end{eqnarray}
where $U_{\alpha i}$ are parameterized by $\theta_{ij}$, $i,j=1,2,3, i<j$, in the usual way \cite{Agashe:2014kda}. From Eq.~(\ref{W_num}), it is easy to see that $W_{\alpha i}\sim U_{\alpha i}$ for the mostly active states ($i=1,2,3$), while the top-left $(3\times 3)$-submatrix of $W$ is not quite unitary. The slow decrease of $S$ as the KK-number increases can be readily observed. It is also easy to see that the effects of the mass eigenstates $7,8,9$, proportional to $|U|^2$ are suppressed relative to those of states $4,5,6$ by a factor of four. One can quickly check that all are significantly smaller than $|U_{e3}|^2$ ($|0.14U_{e1}|^2\sim 0.01$ is the largest $|U_{\alpha i}|$ for $i=7,8,9$ in Eq.~(\ref{W_num})). Furthermore, the oscillation frequencies associated to these states are also four times larger than those from the first KK mode and, for the $R_{\rm ED}$ values of interest, their effects always average out at long-baseline experiments like DUNE. For all these reasons, one set of KK modes is, for DUNE neutrino energies and LED parameters of interest, a good proxy for the LED scenario. As mentioned earlier, all results discussed henceforth were computed including the effects of two KK modes (hence a 3+6 model).

When simulating data consistent with the LED hypothesis, we have to include input values for the $\theta_{ij}$ parameters. When doing that, we try to emulate as well as possible the current best-fit values, which we take to represent the existing neutrino data. In order to do that, we assume that the information that the current data provide for the three-neutrino mixing matrix elements $U_{\alpha i}$ applies to $W_{\alpha i}$ for $i=1,2,3$. Hence, the best-fit value for the LED parameter $\sin^2\theta_{13}$, for example, is not identical to that of the three-neutrino parameter $\sin^2\theta_{13}=0.0219$  \cite{Agashe:2014kda}. They are, however, similar and related. For $R^{-1}_{\rm ED} = 0.38$~eV, $m_0=5\times 10^{-2}$~eV, and the NH, the best fit value for $(\sin^2\theta_{13})_{\rm LED}=0.0219/0.94^2=0.025$ (see Eq.~(\ref{W_num})). This recipe cannot be followed exactly, so we decide on the best-fit, input values for the LED $\theta_{ij}$ parameters by equating the $|W_{e2}|, |W_{e3}|, |W_{\mu 3}|$ to the best-fit values of $|U_{e2}|, |U_{e3}|, |U_{\mu 3}|$ obtained in the three-neutrino framework.

\begin{figure}[t]
	\begin{center}
	\includegraphics[width=1\linewidth]{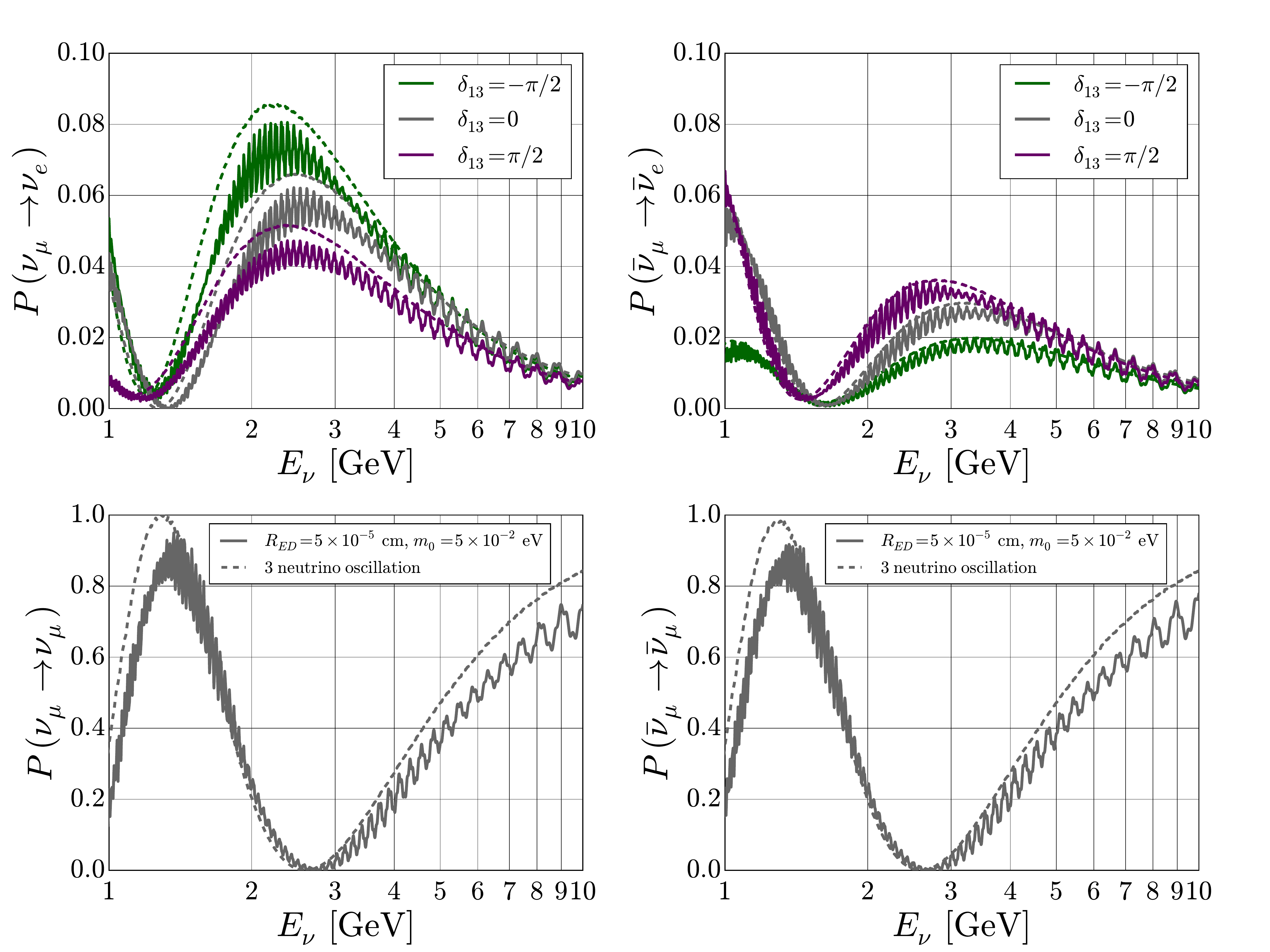}
	\caption{Oscillation probabilities assuming a three-neutrino framework (dashed) and an LED hypothesis with $m_0 = 5\times 10^{-2}$ eV and $R^{-1}_{\rm {ED}} = 0.38$~eV ($R_{\rm {ED}} = 5\times 10^{-5}$ cm), for the normal neutrino mass hierarchy,  $\Delta m_{13}^2 > 0$. The values of the other oscillation parameters are tabulated in Table~\ref{PDGTable}, see text for details. The top row displays appearance probabilities $P(\nu_\mu \to \nu_e)$ (left) and $P(\bar{\nu}_\mu \to \bar{\nu}_e)$ (right), and has curves shown for $\delta_{13} = -\pi/2$ (green), $\delta_{13} = 0$ (gray), and $\delta_{13} = \pi/2$ (purple). The bottom row displays disappearance probabilities $P(\nu_\mu \to \nu_\mu)$ (left) and $P(\bar{\nu}_\mu \to \bar{\nu}_\mu)$ (right).}
	\label{probability}
	\end{center}
\end{figure}

To understand the effect of the LED parameters on the oscillation of neutrinos, we show in Fig.~\ref{probability} the probabilities of $\nu_{\mu}\to\nu_e$ (top-left) and $\bar{\nu}_\mu\to\bar{\nu}_e$ (top-right) as well as the survival probabilities of $\nu_\mu$ (bottom-left) and $\bar{\nu}_\mu$ (bottom-right) in the energy range of DUNE for the three-neutrino scheme and the LED formalism with dashed and solid curves, respectively. In all the panels we have fixed the parameter $\Delta m^2_{j1}$, $j=2,3$ and $\theta_{ij}$, $i,j=1,2,3, i<j$, to the best fit values reported in Ref.~\cite{Agashe:2014kda} (see also Table~\ref{PDGTable}), for 3 different values of $\delta_{13}$. For the LED hypothesis, we further choose $m_0 = 5\times 10^{-2}$ eV and $R^{-1}_{\rm ED} = (5\times 10^{-5}~{\rm cm})^{-1} = 0.38$~eV. We see that for fixed values of $\theta_{ij}$, the oscillation probabilities in the LED case are suppressed with respect to the three-flavor scenario, as discussed above. This effect can be partially remedied by increasing the values of the LED $\theta_{ij}$ parameters. Fig.~\ref{probability} also clearly depicts the fast oscillations associated to the presence of the KK modes.

\section{Excluding the LED Hypothesis}
\label{Sensitivity}
In this section we investigate the sensitivity of DUNE to the model described in Sec.~\ref{Formalism}. We assume, as laid out in \cite{Acciarri:2016crz,Acciarri:2015uup}, that DUNE is comprised of a 34-kiloton liquid argon detector located 1300 km from the neutrino source at Fermilab. The neutrino or antineutrino beam is produced by directing a 1.2 MW beam of protons onto a fixed target. We use the neutrino fluxes and reconstruction efficiencies reported in Ref.~\cite{Adams:2013qkq}\footnote{These are similar but not identical to the ones discussed in Ref.~\cite{Acciarri:2015uup}. Ref.~\cite{Acciarri:2015uup} reports updated reconstruction efficiencies which lead to reduced neutral current backgrounds for the appearance channels. In this light, our results can be viewed as somewhat conservative.} to calculate event yields, as well as the neutrino-nucleon cross-sections reported in Ref.~\cite{Formaggio:2013kya}. The neutrino energies range from 0.5 GeV to 20.0 GeV with maximum flux at around 3.0 GeV. Events are binned in 0.25~GeV bins from 0.5 GeV to 8.0 GeV, resulting in 30 independent counting measurements for each of the four data samples discussed below. Our analysis thus contains 120 degrees of freedom before subtracting the number of parameters describing any particular hypothesis. We simulate a detector resolution of $\sigma \text{[GeV]} = 0.15/\sqrt{E \text{[GeV]}}$ for electrons and $\sigma \text{[GeV]} = 0.20/\sqrt{E \text{[GeV]}}$ for muons, and assume three years of operation each for the neutrino beam and the antineutrino beam.

When generating data assuming the standard three-neutrino framework, we assume  the best-fit values for the oscillation parameters from Ref.~\cite{Agashe:2014kda}, summarized in Table~\ref{PDGTable}. Since the neutrino mass hierarchy is unknown, we simulate data using either the normal hierarchy (NH) or inverted hierarchy (IH). We assume, however,  that the hierarchy will be known by the time DUNE collects data and therefore analyze the simulated data with the correct hierarchy hypothesis.
\begin{table}[ht]
\centering
\begin{tabular}{| c || c | c | }
\hline
Parameter & Normal Hierarchy & Inverted Hierarchy \\
\hline
$\sin^2\theta_{12}$ & $0.304\pm 0.014$ & $0.304\pm 0.014$ \\
\hline
$\sin^2\theta_{13}$ & $(2.19\pm 0.12)\times 10^{-2}$ & $(2.19\pm 0.12)\times 10^{-2}$ \\
\hline
$\sin^2\theta_{23}$ & $0.514^{+0.055}_{-0.056}$ & $0.511\pm 0.055$ \\
\hline
$\Delta m_{21}^2$ & $(7.53\pm 0.18)\times 10^{-5}$ eV$^2$ & $(7.53\pm 0.18)\times 10^{-5}$ eV$^2$ \\
\hline
$\Delta m_{31}^2$ & $(2.51\pm 0.06)\times 10^{-3}$ eV$^2$ & $-(2.41\pm 0.06)\times 10^{-3}$ eV$^2$ \\
\hline
$|U_{e2}|^2$ & $0.297\pm 0.014$ & $0.297\pm 0.014$ \\
 \hline
\end{tabular}
\caption{Best-fit values of three-neutrino mixing parameters assuming the normal or inverted mass hierarchy. Values come from the 2015 update to Ref.~\cite{Agashe:2014kda}, and the parameter $|U_{e2}|^2$, which is used later in our analysis, is derived from the fits to $\sin^2\theta_{12}$ and $\sin^2\theta_{13}$. While there exist, currently, weak constraints on the $CP$-odd parameter $\delta_{13}$, we work under the assumption that it is unconstrained.}
\label{PDGTable}
\end{table}

\begin{figure}[t]
	\begin{center}
	\includegraphics[width=0.95\linewidth]{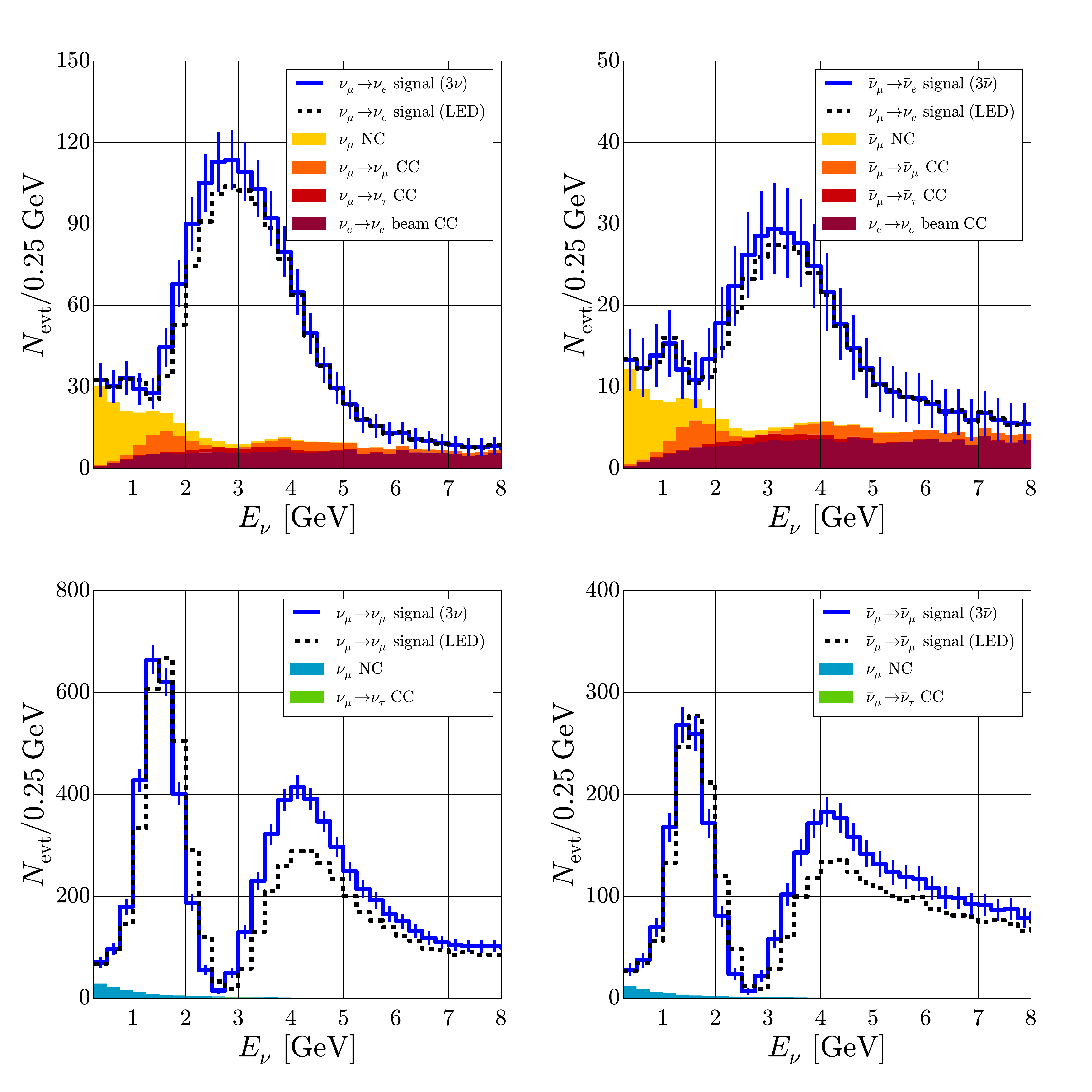}	
	\caption{Expected event yields at DUNE assuming three years of either neutrino-beam mode (left) or antineutrino-beam mode (right). The top row displays $\nu_e$ and $\bar{\nu}_e$ appearance yields and the bottom row displays $\nu_\mu$ and $\bar{\nu}_\mu$ disappearance yields. In each panel, we show the expected yield assuming a three-neutrino hypothesis with parameters from Table~\ref{PDGTable} for the normal hierarchy in blue, with error bars representing statistical uncertainties, and assuming a non-zero LED hypothesis with $m_0 = 5\times 10^{-2}$ eV and $R_{\rm ED}^{-1} = 0.38$~eV in black. The contribution of events associated to opposite-sign muons and electrons is included in the signal. Backgrounds are discussed in the text and shown under the expected signals.}
	\label{fig:4Pane}
	\end{center}
\end{figure}
Fig.~\ref{fig:4Pane} displays expected event yields for neutrino appearance ($P(\nu_\mu\to\nu_e)$, top-left), antineutrino appearance ($P(\bar{\nu}_\mu \to \bar{\nu}_e)$, top-right), neutrino disappearance ($P(\nu_\mu \to \nu_\mu)$, bottom-left), and antineutrino disappearance ($P(\bar{\nu}_\mu\to\bar{\nu}_\mu)$, bottom-right). In each panel, the expected event yield at DUNE is displayed for a three-neutrino hypothesis with parameters from Table~\ref{PDGTable} for the normal hierarchy, $\delta_{13}=0$, and for a non-zero LED hypothesis with all homonymous parameters the same plus $m_0 = 5\times 10^{-2}$ eV and $(R_{\rm ED})^{-1} = 0.38$~eV.\footnote{This is done for illustrative purposes only. The set of LED parameters that best mimics the three-flavor paradigm will have best-fit values of, for example, $\theta_{ij}$, $i,j=1,2,3, i<j$, that are different from the input three-flavor values for $\theta_{ij}$, as discussed earlier.} The dominant backgrounds are neutral-current scattering of muon-neutrinos (``$\nu_\mu$ NC"); charged-current scattering of tau-neutrinos  (``$\nu_\mu \to \nu_\tau$ CC"); neutral-current scattering of unoscillated muon-type neutrinos  (``$\nu_\mu \to \nu_\mu$ NC"); and charged-current scattering of unoscillated, contaminant electron-type neutrinos (``$\nu_e \to \nu_e$ CC"). The rates of these processes are estimated from Ref.~\cite{Adams:2013qkq}, and are not recalculated in our analyses for different hypotheses, as 1\% signal and 5\% background normalization uncertainties overwhelm any noticeable effects.  

\begin{figure}[t]
	\begin{center}
	\subfigure[]{\label{subfig:NHSens}\includegraphics[width=0.49\linewidth]{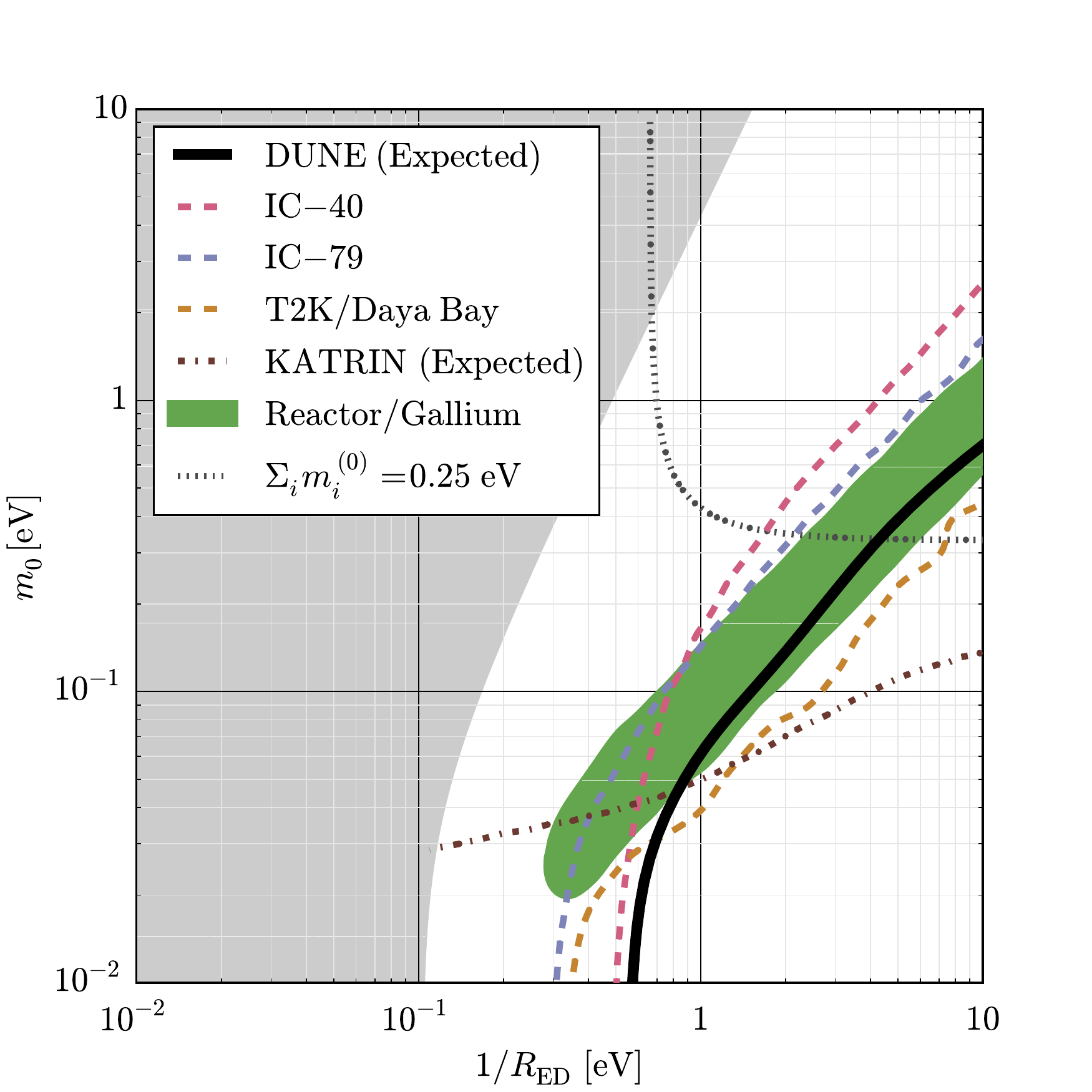}}
	\subfigure[]{\label{subfig:IHSens}\includegraphics[width=0.49\linewidth]{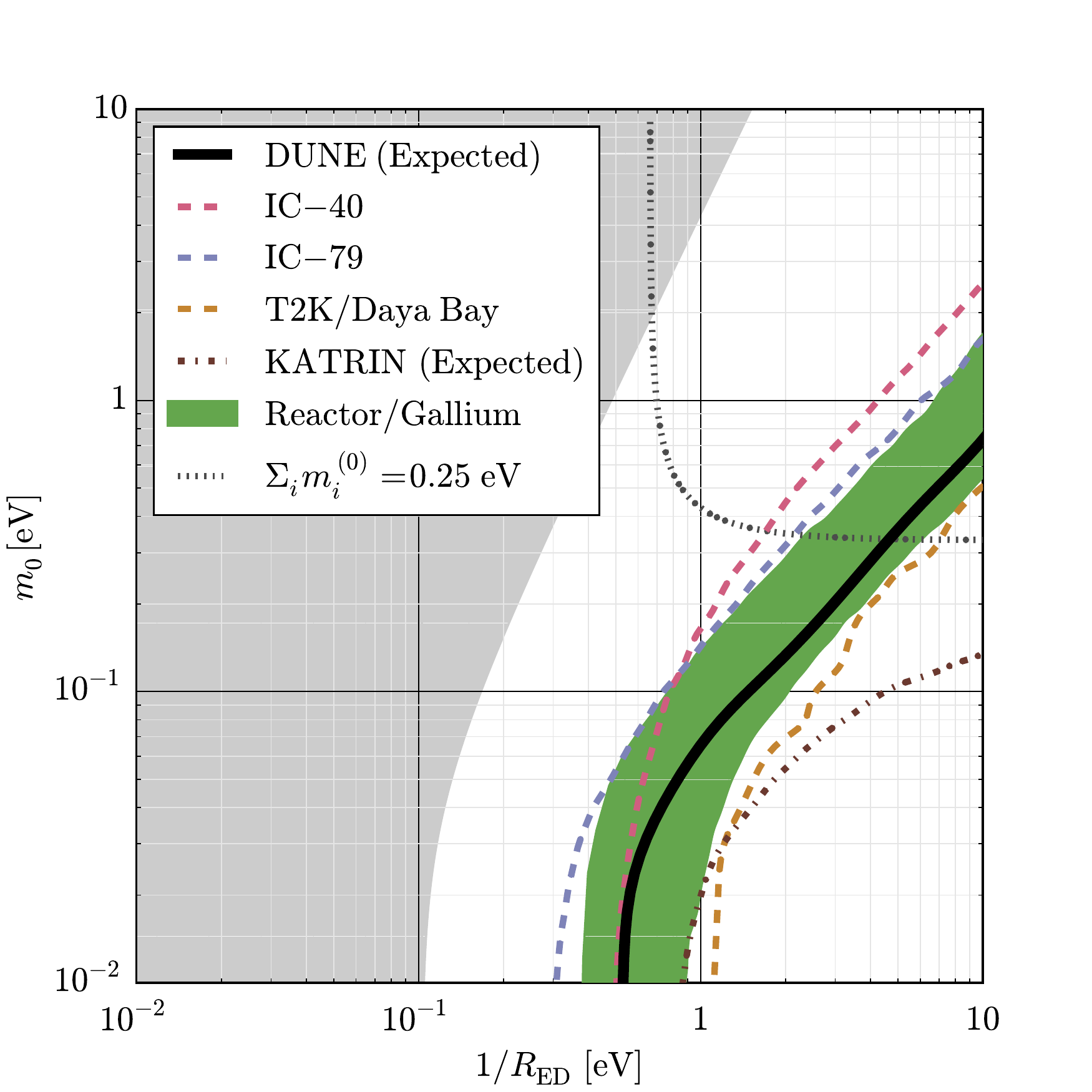}}
	\caption{Exclusion limits in the $R_{\rm ED}^{-1}$--$m_{0}$ plane, assuming either (a) a normal hierarchy or (b) an inverted hierarchy of neutrino masses. The exclusion regions are to the top-left of the relevant curves. Shown are the 95\% CL lines from DUNE (black), IceCube-40 (mauve) and Ice-Cube79 (blue) \cite{Esmaili:2014esa}, and a combined analysis of T2K and Daya Bay (gold)~\cite{DiIura:2014csa}. We also include the 90\% CL line from sensitivity analysis of KATRIN (burgundy)~\cite{BastoGonzalez:2012me}. The shaded green regions are preferred at 95\% CL by the reactor anomaly seen in reactor and Gallium experiments~\cite{Machado:2011kt}. The gray shaded regions are excluded by the measurements of $\Delta m^2_{i1}$, as explained in the text. The dotted gray lines are curves along which $\sum_i m^{(0)}_i = 0.25$ eV. Higher values of $\sum_i m^{(0)}_i$ correspond to the regions above and to the right of the dotted gray lines.}
	\label{fig:Sens}
	\end{center}
\end{figure}

We analyze pseudodata simulated under the standard three-neutrino framework plus $\delta_{13}=0$ with the LED hypothesis. The resulting 95\% confidence level (CL) limit in the $R_{\rm ED}^{-1}$--$m_{0}$ plane is shown in black in Fig.~\ref{subfig:NHSens} for the NH and in Fig.~\ref{subfig:IHSens} for the IH. In the analysis, following Refs.~\cite{Berryman:2015nua,deGouvea:2015ndi}, we include priors on the solar parameters in order to take constraints from solar and KamLAND data into account. More concretely, we add Gaussian priors on $\Delta m^2_{21}$  using the information in Table~\ref{PDGTable}, and on $|W_{e2}|^2$ using the information for $|U_{e2}|^2$ tabulated in Table~\ref{PDGTable}. In the analysis, we marginalize over all parameters not made explicit in the figures.  We have repeated the analysis for several nontrivial input values of $\delta_{13}$ and find the corresponding exclusion limits to be similar to the ones depicted in Fig.~\ref{fig:Sens}. 

The dashed mauve and blue curves in Fig.~\ref{fig:Sens} show the exclusion limits at 95\% CL from IceCube-40 data and IceCube-79 data, respectively, as calculated in Ref.~\cite{Esmaili:2014esa}. The dashed gold curves are the same for a combined analysis of T2K and Daya Bay performed in Ref.~\cite{DiIura:2014csa}. The green regions are preferred at 95\% CL by short-baseline oscillation experiments according to analysis published in Ref.~\cite{Machado:2011kt}. All these curves have, to zeroth order, the same shape as the exclusion curve we obtain for DUNE. This happens because the ratio of $m_0$ and $R^{-1}_{\rm ED}$, when small, can be mapped into an effective mixing angle which governs most oscillation phenomena, as discussed in Ref.~\cite{Esmaili:2014esa}. 

The dot-dashed burgundy curves in Fig.~\ref{fig:Sens} show the expected 90\% CL exclusion limit of the $\beta$-decay experiment KATRIN, estimated in Ref.~\cite{BastoGonzalez:2012me}. The dependence on $m_0$ and on $R_{\rm ED}^{-1}$ is more complicated for $\beta$-decay experiments than for oscillation experiments as the former rely on kinematic information from the electrons emitted in the decay.

The gray shaded regions are excluded on the basis of the mass-squared differences $\Delta m^2_{21}$ and $\Delta m^2_{31}$. As discussed in Sec.~\ref{Formalism}, $\Delta m^2_{i1},~i=2, 3$ characterize the differences between the lowest-lying\footnote{Observed oscillations cannot be due to mixing among mass states from different KK modes. The mixing with the other low-lying state(s) would be large enough to produce a deviation from the three-standard-paradigm that is inconsistent with existing neutrino oscillation data.} physical masses-squared differences, $[(\lambda^{(0)}_i)^2 -(\lambda^{(0)}_1)^2] / R_{\rm ED}^2$. The transcendental equation Eq.~(\ref{trans}) can only be satisfied if $0 < \lambda^{(0)}_i < 0.5$. Therefore, a point in the $R_{\rm ED}^{-1}$--$m_{0}$ plane is only physical if all $\lambda^{(0)}_i$ implied by $\Delta m^2_{21}$ and $\Delta m^2_{31}$ meet this requirement; the unphysical points define the gray shaded regions~\cite{BastoGonzalez:2012me}.

The dotted gray lines are curves along which the sum of the masses of the three mostly active eigenstates, $\sum_i m^{(0)}_i$, is 0.25 eV. This value is roughly the same as the current upper bound on the sum of the neutrino masses from PLANCK \cite{Ade:2015xua}. A proper analysis of the cosmology of the LED framework is outside the scope of this work.  However, we believe the dotted gray lines capture the spirit of potential cosmological bounds in the $R_{\rm ED}^{-1}$--$m_{0}$ plane, especially if one allows for possible extensions of the LED scenario under consideration here.


\setcounter{footnote}{0}

\section{Measuring LED parameters}
\label{Measuring}

\begin{figure}[t]
\begin{center}
\subfigure[]{\label{subfig:RxM1_NH}\includegraphics[width=0.49\linewidth]{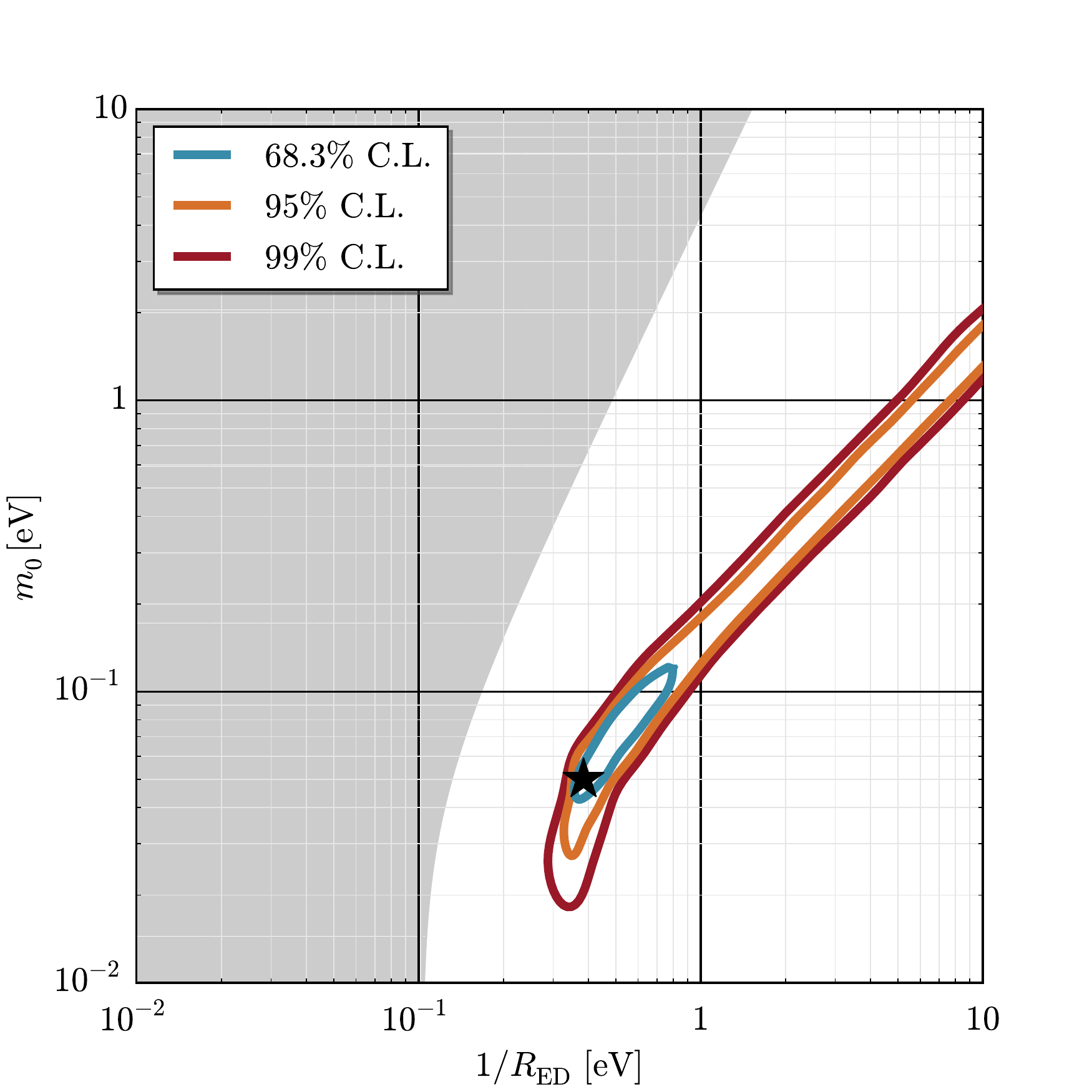}}
\subfigure[]{\label{subfig:RxM1_IH}\includegraphics[width=0.49\linewidth]{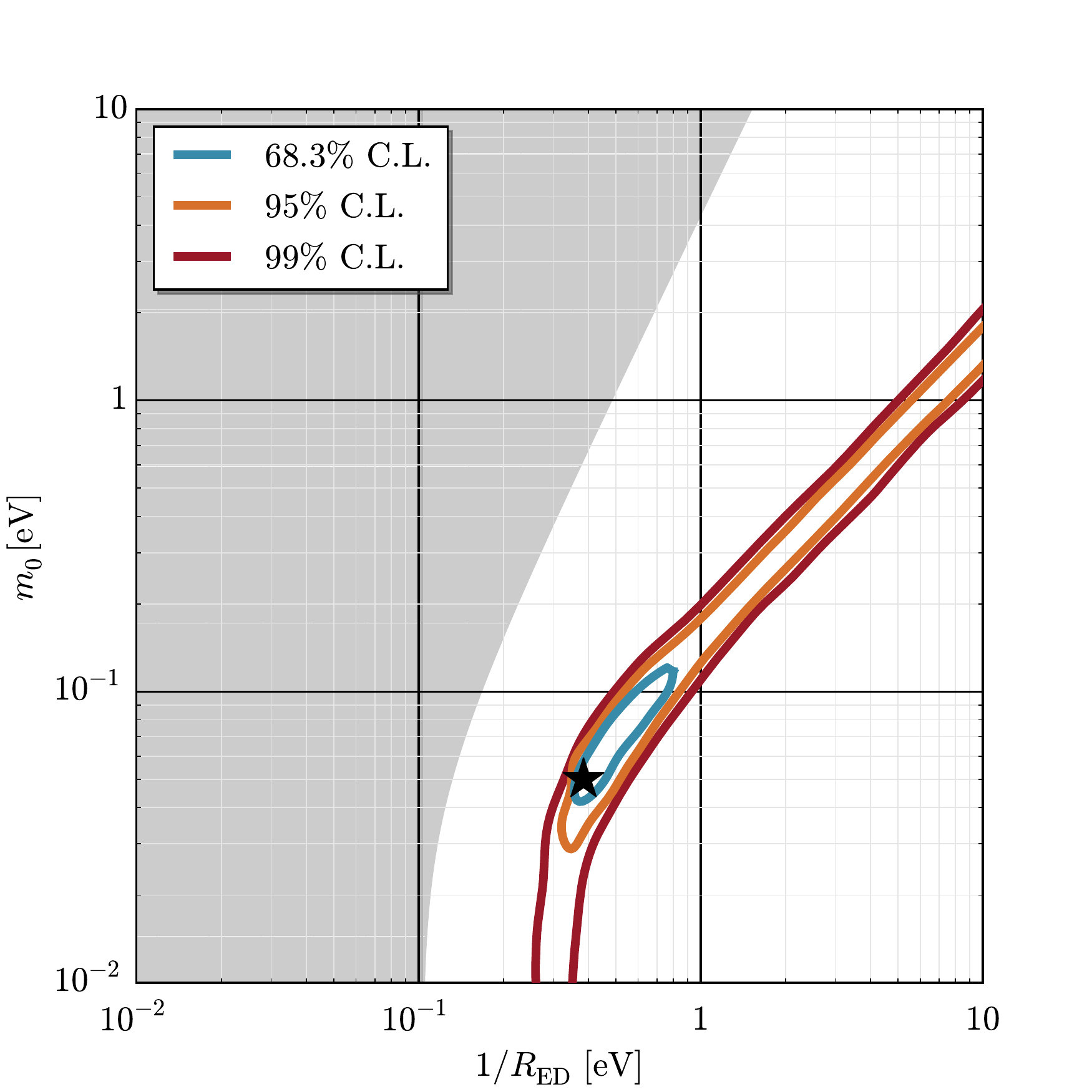}}
\caption{Expected sensitivity to a non-zero set of LED parameters as measured by DUNE, assuming three years each of neutrino and antineutrino data collection. Fig.~\ref{subfig:RxM1_NH} assumes the normal mass hierarchy (NH) and Fig.~\ref{subfig:RxM1_IH} assumes the inverted mass hierachy (IH). The LED parameters assumed here are $m_0 = 5\times 10^{-2}$ eV and $R^{-1}_{\rm ED} = 0.38$~eV, while $\delta_{13}=\pi/3$. The input values of $\Delta m^2_{i1}$, $i=1,2$ are in Table~\ref{PDGTable}. The input values for the mixing angles are, for the NH, $\sin^2\theta_{12}=0.322$, $\sin^2\theta_{13}=0.0247$, $\sin^2\theta_{23}=0.581$, and, for the IH, $\sin^2\theta_{12}=0.343$, $\sin^2\theta_{13}=0.0231$, $\sin^2\theta_{23}=0.541$.}
\label{fig:RxM1Sens}
\end{center}
\end{figure}

In this section we simulate data consistent with the  LED hypothesis and investigate how well DUNE is capable of measuring the new-physics parameters $m_0$ and $R^{-1}_{\rm ED}$ in tandem with the other oscillation parameters, introduced in Sec.~\ref{Formalism}. As input, we use the values for $\Delta m^2_{i1}$, $i=2,3$ tabulated in Table~\ref{PDGTable}, for the normal and inverted hierarchies, and choose $\delta_{13}=\pi/3$, $m_0 = 5\times 10^{-2}$ eV, and $R^{-1}_{\rm ED} = 0.38$~eV. We choose these values to be in the region excluded by DUNE shown in Fig.~\ref{fig:Sens}. As discussed earlier, we choose $\theta_{ij}$, $i,j=1,2,3, i<j$, such that $|W_{e2}|,|W_{e3}|,|W_{\mu 3}|$ agree with the best-fit values of $|U_{e2}|,|U_{e3}|,|U_{\mu 3}|$ under the three-flavor hypothesis.
As discussed in Sec.~\ref{Sensitivity}, we add Gaussian priors for the solar parameters, identified here as $\Delta m^2_{21}$ and $|W_{e2}|^2$. The results of these fits are depicted in Fig.~\ref{fig:RxM1Sens}. In the analysis, we marginalize over all parameters not made explicit in the figures. 

Fig.~\ref{fig:RxM1Sens} reveals that, at least at 99\% CL, a lower bound on $R_{\rm ED}^{-1}$ can be obtained in both the normal and inverted hierarchy scenarios, while a lower bound on $m_{0}$ can be set at least at 95\% CL for both mass hierarchies. Additionally, if one were to place an independent bound on different combinations of  neutrino masses (from, e.g., precision measurements of beta-decay spectra), a 99\% CL upper bound on $R_{\rm ED}^{-1}$ (or a lower bound on $R_{\rm ED}$) could be obtained.  

Finally, we have verified that the presence of the LED parameters $m_0,R^{-1}_{\rm ED}$ does not significantly impact the sensitivity with which the standard oscillation parameters are measured (see, e.g., Refs.~\citep{Adams:2013qkq,Berryman:2015nua} for more details). This includes the $CP$-odd parameter $\delta_{13}$. We have also checked that this result does not depend strongly on the input value of $\delta_{13}$. 


\section{Differentiating New Physics Scenarios}
\label{Diagnosis}

In this section we address the capabilities of DUNE to identify whether there is physics beyond the three-flavor paradigm and identify the nature of the new physics, assuming new physics is indeed present. To that effect, in Sec.~\ref{subsec:3NuFit}, we first simulate data consistent with the LED hypothesis, as we did in Sec.~\ref{Measuring}, and try to fit the data with the three-neutrino hypothesis. We then ask whether it is possible to differentiate the LED hypothesis from other new physics scenarios. In particular, we compare the LED hypothesis with that of a fourth neutrino mass eigenstate.  In Sec.~\ref{subsec:4NuFit}, we address whether a four-neutrino model can mimic the LED hypothesis, while in Sec.~\ref{subsec:LEDFit} we ask the opposite question: can the LED hypothesis mimic generic four-neutrino models?

\subsection{Three-Neutrino Fit to the LED Scenario}
\label{subsec:3NuFit}
In order to gauge whether DUNE can rule out the standard paradigm, we simulate data assuming the LED hypothesis is correct, exactly as described in Sec.~\ref{Measuring}, and attempt to fit the data assuming the standard, three-neutrino paradigm. The fit is performed for two simulated data sets, consistent with the normal and inverted hierarchies respectively. In order to gauge the quality of the fit, we  calculate the minimum of the $\chi^2$ function, $\chi^2_{\text{min}}$, and compare it to the number of degrees of freedom, dof.  We define an equivalent $n\sigma$ discrepancy between the data and hypothesis assuming a $\chi^2$ distribution function with dof degrees of freedom. In the fits, we include the Gaussian priors on $|U_{e2}|^2$ and $\Delta m_{21}^2$, as discussed in the previous sections (see also \cite{Berryman:2015nua,deGouvea:2015ndi}).

For the normal hierarchy, the result of the fit is $\chi^2_{\text{min}} /$ dof = $210/114$, or a $5.3\sigma$ discrepancy -- a very poor fit. For the inverted hierarchy, the fit is $\chi^2_{\text{min}} /$ dof = $208/114$, or a $5.2\sigma$ discrepancy -- also a very poor fit. These results are, of course, not surprising. According to Fig.~\ref{fig:Sens}, the input values of $R^{-1}_{\rm ED}$ and $m_{0}$ are far inside the region of LED parameter space DUNE can exclude at 95\% CL.

\subsection{Four-Neutrino Fit to the LED Scenario}
\label{subsec:4NuFit}
If data are consistent with the LED hypothesis so the standard paradigm is ruled out, it is not obvious that DUNE can establish that there are extra dimensions. The LED hypothesis is identical to a $3+3N$ active-plus-sterile-neutrinos scenario for large enough $N$. In fact, we argued in the Sec.~\ref{Formalism} that, for the values of the parameters of relevance here, $N=1$ is already a good approximation to the LED model. Here, we attempt to fit the simulated LED model to a four-neutrino hypothesis, using the framework described in Ref.~\cite{Berryman:2015nua}.\footnote{We denote the six mixing angles in a four-neutrino hypothesis as $\phi_{ij}$ ($i,j=1,2,3,4, i<j$) to emphasize that they are not equivalent to the $\theta_{ij}$ of a three-neutrino hypothesis. The $CP$-violating phase $\eta_1$ is equivalent to $\delta_{13}$, and the new phases $\eta_2$ and $\eta_3$ contribute in the appearance channel in the combination $\eta_s \equiv \eta_2 - \eta_3$.} While four neutrinos is less than the six neutrinos that are known to be a good approximation to the LED hypothesis, there is reason to suspect that, at DUNE and given the values of $m_0$ and $R^{-1}_{\rm ED}$ of interest, the four-neutrino hypothesis is also a good approximation to the LED model. The reason is as follows. At the DUNE baseline and given DUNE neutrino energies, oscillation effects associated to the KK modes average out. The same effect can be mimicked by a 3+1 scenario in the limit where the new mass-squared difference is large. The map between the 3+1 and the LED scenario is not completely straightforward, but there are enough relevant degrees of freedom in the 3+1 model to accommodate all LED effects assuming there are no new resolvable mass-squared differences.\footnote{Seven, $\phi_{12},\phi_{13}, \phi_{23},\phi_{14},\phi_{24},\eta_1,\eta_{s}$ in the 3+1 case, compared to six, $\theta_{12}, \theta_{13},\theta_{23},\delta_1,m_0,R_{\rm ED}$, in the LED case.}

For both the NH and IH, we find a good fit (i.e., $\chi^2_{\text{min}} \simeq$ dof). The results of these fits, one for each hierarchy hypothesis, are summarized in Table~\ref{4NuFitTable}. For both hierarchies, the four-neutrino hypothesis favors $\Delta m_{41}^2 > 0.1$ eV$^2$, the range in which oscillations associated with the extra neutrino average out for the energies of interest at DUNE. For this reason, we expect little sensitivity to the new, potentially observable, $CP$-violating phase $\eta_s \equiv \eta_2 - \eta_3$. 
\begin{table}[ht]
\centering
\begin{tabular}{| c || c | c | }
\hline
Parameter & Normal Hierarchy (NH) & Inverted Hierarchy (IH) \\
\hline
$\sin^2\phi_{12}$ & $0.311^{+0.028}_{-0.033}$ & $0.287^{+0.051}_{-0.010}$ \\
\hline
$\sin^2\phi_{13}$ & $\left(2.28^{+0.60}_{-0.40}\right)\times 10^{-2}$ & $\left(1.95^{+0.73}_{-0.31}\right)\times 10^{-2}$ \\
\hline
$\sin^2\phi_{23}$ & $0.523^{+0.030}_{-0.042}$ & $0.532^{+0.022}_{-0.056}$ \\
\hline
$\sin^2\phi_{14}$ & $\left(6.20^{+16.13}_{-6.20}\right)\times 10^{-3}$ & $\left(9.06^{+13.27}_{-9.06}\right)\times 10^{-3}$ \\
\hline
$\sin^2\phi_{24}$ & $\left(5.65^{+1.15}_{-1.31}\right)\times 10^{-2}$ & $\left(6.76^{+0.36}_{-2.41}\right)\times 10^{-2}$ \\
\hline
$\sin^2\phi_{34}^\star$ & $0$ & $0$ \\
\hline
$\Delta m_{21}^2$ & $\left(7.50^{+0.45}_{-0.33}\right)\times 10^{-5}$ eV$^2$ & $\left(7.68^{+0.27}_{-0.51}\right)\times 10^{-5}$ eV$^2$ \\
\hline
$\Delta m_{31}^2$ & $\left(2.69^{+0.02}_{-0.03}\right)\times 10^{-3}$ eV$^2$ & $\left(-2.58^{+0.03}_{-0.04}\right)\times 10^{-3}$ eV$^2$ \\
\hline
$\Delta m_{41}^2$ & $\left(0.57^{+1.42}_{-0.37}\right)$ eV$^2$ & $\left(0.56^{+1.44}_{-0.36}\right)$ eV$^2$ \\
\hline
$\eta_1$ & $\left(0.54^{+0.04}_{-0.36}\right)\pi$ & $\left(0.38^{+0.16}_{-0.1320}\right)\pi$ \\
\hline
$\eta_s \equiv \eta_2 - \eta_3$ & $\left(-0.03^{+1.03}_{-0.97}\right)\pi$ & $\left(-0.04^{+1.04}_{-0.96}\right)\pi$ \\
\hline
\end{tabular}
\caption{Results of four-neutrino fits to data generated according to the LED Hypotheses discussed in Sec.~\ref{Measuring}. Best-fit values are the result of a 10-dimensional minimization, while quoted 95\% CL ranges are from the marginalized one-dimensional resulting $\chi^2$ distributions for each parameter. The star on $\sin^2\phi_{34}$ is a reminder that we are not including $\nu_{\tau}$-appearance information and hence have no sensitivity to $\sin^2\phi_{34}$. For this reason, we fix it to zero. See Ref.~\cite{Berryman:2015nua} for more information.}
\label{4NuFitTable}
\end{table}

\begin{figure}[t]
\begin{center}
	\includegraphics[width=\linewidth]{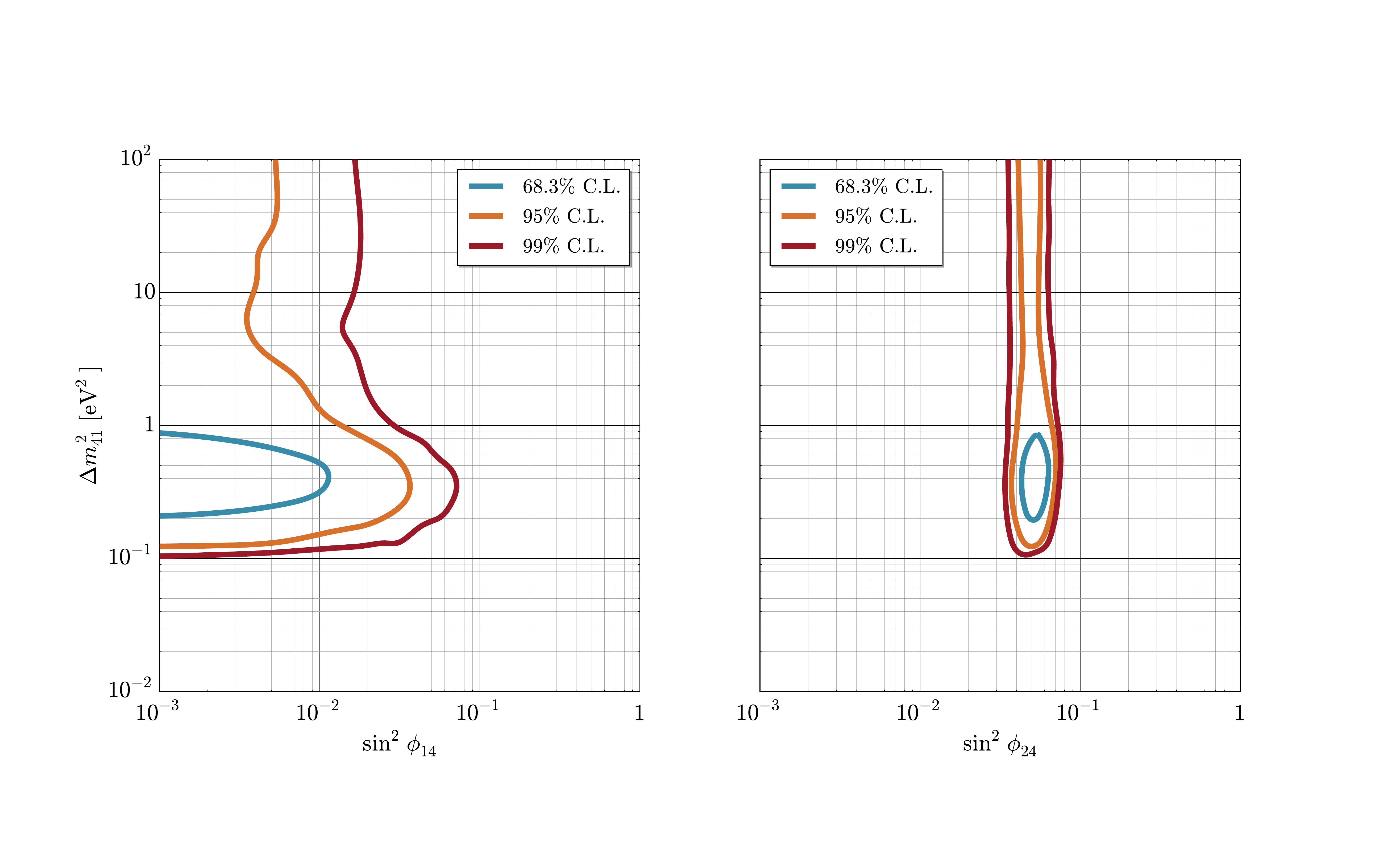}
	\caption{Results of a four-neutrino fit to data generated assuming an LED hypothesis with $m_0 = 5\times 10^{-2}$ eV and $R^{-1}_{\rm ED} = 0.38$~eV assuming the normal hierarchy. Contours shown are 68.3\% (blue), 95\% (orange), and 99\% (red) CL. All unseen parameters are marginalized over.}
	\label{fig:4NuFittoLED}
\end{center}
\end{figure}

Fig.~\ref{fig:4NuFittoLED} displays the result of the fit performed assuming the normal hierarchy in the $\sin^2\phi_{14}$ - $\Delta m_{41}^2$ and $\sin^2\phi_{24}$ - $\Delta m_{41}^2$ planes. We find a qualitatively similar result when performing the fit assuming the neutrino mass hierarchy is inverted.  Note that the data are consistent with  $\sin^2\phi_{14} = 0$ at 68.3\% CL, but $\sin^2\phi_{24} = 0$ is excluded at more than 99\% CL. On the other hand, while it is possible to establish that the new oscillation frequency is large ($\Delta m^2_{41}>0.1$~eV$^2$ at a high confidence level), it is not possible to place an upper bound on the new mass-squared difference. 

\subsection{LED Fit to Four-Neutrino Scenarios}
\label{subsec:LEDFit}
Here, we generate data assuming four neutrinos exist, and attempt to fit this simulated data under the LED hypothesis. While it is easy to show that the LED hypothesis, under the circumstances of interest, can be mimicked by a four-neutrino scenario, the converse is by no means obvious. In the LED hypothesis, the elements of the (infinitely large) neutrino mixing matrix are all related and can be uniquely determined once a handful of parameters are fixed, as described in Sec.~\ref{Formalism}. This means that the LED hypothesis can only perfectly mimic a four-neutrino scenario if the mixing angles and $CP$-odd parameters are related in nontrivial ways. In summary, at least at the oscillation probability level, a generic four-neutrino scenario cannot be mimicked by the LED hypothesis. 

We pursue the issue by perturbing around the best-fit solutions discussed in the previous subsection and tabulated in Table~\ref{4NuFitTable}. First, we generate data assuming the four-neutrino parameters listed in Table~\ref{4NuFitTable}. In this case, for both the normal and inverted hierarchies, we find that the LED hypothesis generates a good ($\chi^2_{\text{min}} \simeq$ dof) fit, with $m_0/(R_{\rm ED})^{-1} \simeq 0.13$, which is what we expect given the original LED hypothesis we assumed in Sec.~\ref{Measuring}.

Next, we generate data assuming the four-neutrino parameters listed in Table~\ref{4NuFitTable} but with $\Delta m_{41}^2 = 10^{-2}$ eV$^2$, a value studied more in-depth in Ref.~\cite{Berryman:2015nua}. For this value of $\Delta m_{41}^2$, we expect the new oscillations due to the fourth neutrino to be relevant for the energies of interest at DUNE. In this case, for the normal hierarchy, we obtain a fit that has $\chi^2_{\text{min}} /$ dof $= 349/112$, which corresponds to a discrepancy larger than  $8\sigma$ -- a very poor fit. For the inverted hierarchy, the fit has $\chi^2_{\text{min}} /$ dof $= 402/112$, corresponding to a larger than $8\sigma$ discrepancy  -- also a very poor fit. In either case, DUNE would be able to rule out both the three-flavor hypothesis and the LED hypothesis, while the four-neutrino hypothesis would provide an excellent fit to the data. 

We repeat the exercise, this time assuming the input values of all the four-neutrino parameters are those listed in Table~\ref{4NuFitTable}, except for the new mixing angles. If the input values of $\sin^2\phi_{14}$ and $\sin^2\phi_{24}$ are 0.1 and 0.01, respectively, the LED hypothesis also fails to fit the 3+1 scenario, for either mass hierarchy: $\chi^2_{\text{min}} /$ dof $= 213/112$ ($6.0\sigma$) for the NH, $\chi^2_{\text{min}} /$ dof $= 241/112$ ($6.7\sigma$) for the IH. In summary, at DUNE, the LED hypothesis can always be mimicked by the 3+1 scenario, but the converse is, by no means, generically true. 


\section{CONCLUSIONS}
\label{Concluding}

The long-baseline Deep Underground Neutrino Experiment (DUNE)~\cite{Adams:2013qkq} has been proposed to address several outstanding issues in neutrino physics, including the search for new sources of $CP$-invariance violation and precision tests of the validity of the standard three-massive-neutrinos paradigm. In this work, we addressed the ability of DUNE to probe large-extra-dimension (LED) models. These are scenarios where the smallness of neutrino masses is, at least partially, attributed to the existence of one extra compactified dimension of space which is accessible to the right-handed neutrino fields but inaccessible to all fields charged under the standard model gauge group.  From a four-dimensional point of view, the Kaluza-Klein (KK) expansion of the right-handed neutrinos translates into towers of massive sterile neutrino states, with masses inversely proportional to the size $R_{\rm ED}$ of the extra dimension. 

We discussed in some detail the phenomenon of neutrino oscillations at long-baseline experiments in a five-dimensional LED model. We argued that the LED model, for all practical purposes, maps into a $3+3N$-neutrino scenario, and that modest values of $N$, $N=1$ or $N=2$ capture the details of the LED effects at long-baseline oscillations experiments. Nonetheless, we emphasized that the LED model does not map into a generic $3+3N$ model. Instead, the number of new independent mixing parameters is small -- six, including four that can be interpreted, to leading order, as the familiar three-neutrino mixing parameters $\theta_{12}, \theta_{23},\theta_{13},\delta_{13}$. Furthermore, we highlighted the fact that in LED models, there are no new $CP$-invariance violting parameters; the only source is the $CP$-odd phase $\delta_{13}$, which, to zeroth order, plays the same role in the three-neutrino scenario.

We investigated the sensitivity of DUNE to the LED framework. Assuming future DUNE data are consistent with the three-neutrino paradigm (assuming three and three years of operation in neutrino and antineutrino modes, respectively), the LED paradigm can be excluded at $95\%$ CL if $R_{\rm{ED}}^{-1}\le0.54$~eV ($R_{\rm{ED}}^{-1}\le0.48$~eV) assuming a normal (inverted) hierarchy for the mostly active neutrinos. More stringent limits are obtained if $m_0$, related to the mass of the mostly active states, turns out to be large ($m_0\gtrsim 0.01$~eV). The reach of DUNE is compared to that of existing and future probes in Fig.~\ref{fig:Sens}.

We also investigated whether DUNE can measure the new physics parameters if its data turn out to be consistent with the LED model. We found that there are values of $m_0$ and $R_{\rm{ED}}^{-1}$ for which DUNE can establish, at least at the 68\% CL, that $m_0$ is not zero and that the extra-dimension has a finite size. One concrete example is depicted in Fig.~\ref{fig:RxM1Sens}. 

Finally, we explored whether, assuming DUNE data are inconsistent with the three-neutrino paradigm, whether they can reveal the nature of the new physics. We found that data consistent with LED models are inconsistent with the three-neutrino model if the new physics effects are strong enough. Nonetheless, we also found that, as far as DUNE is concerned, there are four-neutrino scenarios which mimic the LED model very effectively. We showed, however, that the converse is not true. If DUNE data are consistent with a four-neutrino scenario, it is likely that the data cannot be explained by an LED scenario. In a nutshell, the LED model, in spite of the fact that it contains an infinite number of new neutrino states, has fewer relevant free parameters than a generic four-neutrino model. 

The key distinguishing features of LED models are the existence of several sterile neutrinos with hierarchical masses (the new masses are, roughly, $R_{\rm ED}^{-1}, 2R_{\rm ED}^{-1}, 3R_{\rm ED}^{-1}, \ldots$) and strongly correlated elements of the infinite mixing matrix ($\alpha 4$ elements proportional to $\alpha 1$ elements, $\alpha 5$ elements proportional to $\alpha 2$ elements, etc, for all $\alpha=e,\mu,\tau$). Both are very difficult to establish experimentally in long-baseline experiments because, in those experiments, the effects of the new oscillation frequencies average out. On the other hand, once new physics effects in $\nu_{\mu}$ disappearance and $\nu_{\mu}\to\nu_e$ appearance are established, the LED hypothesis translates into very concrete predictions for all other oscillation channels, including $\nu_{\mu}\to\nu_{\tau}$ appearance. This is not the case of a generic $3+1$-scenario, where the new-physics effects in the $\nu_{\tau}$-appearance channel cannot be constrained by precision measurements of $\nu_{\mu}$-disappearance and $\nu_e$-appearance.


\section{Acknowledgments}
AdG and ZT thank the hospitality of the Mainz Institute for Theoretical Physics and the ``Crossroads of Neutrino Physics Worshop,'' where this work was initiated. ZT thanks the useful discussions with Arman Esmaili. The work of JMB, AdG, and KJK  is supported in part by the United States Department of Energy grant \#DE-FG02-91ER40684. OLGP thanks the support of FAPESP funding grant 2012/16389-1. The work of ZT is supported by CNPq funding grant 400527/2015-4. \\

 \bibliography{LED_DUNE}
\end{document}